\documentclass[12pt]{article}
\usepackage{amsmath,amsfonts}
\newtheorem{Def}{Definition}

\newtheorem{Remark}{Remark}

\newtheorem{theorem}{Theorem}

\def\sqr#1#2{{\vcenter{\vbox{\hrule height.#2pt
               \hbox{\vrule width.#2pt height#1pt \kern#1pt
                      \vrule width.#2pt}
                      \hrule height.#2pt}}}}

\def\qed{\hbox{\hskip 6pt\vrule width6pt height7pt depth1pt \hskip1pt}}
\def\R{\real}
\def\C{\complex}
\def\nd{\noindent}
\def\real{{\rm I\kern-.2em R}}
\def\complex{\kern.1em{\raise.47ex\hbox{ $\scriptscriptstyle
|$}}\kern-.40em{\rm C}}
\def\be{\begin{equation}}
\def\ee{\end{equation}}
\def\bearray{\begin{eqnarray}}
\def\eearray{\end{eqnarray}}
\def\bra{\langle}
\def\ket{\rangle}

\title{ Fredholm Indices and the Phase Diagram of  Quantum Hall Systems}
\author{J.~E.~Avron and L.~Sadun\footnote{On leave from the
Department of Mathematics,
University of Texas, Austin, TX 78712 USA}
\\ Department of Physics, Technion, 32000 Haifa, Israel}
\begin{document}
\maketitle
\begin{abstract}
The quantized Hall conductance  in a plateau is related to the
index of a Fredholm operator.  In this paper we describe the generic ``phase
diagram'' of Fredholm indices associated with bounded and Toeplitz
operators. We discuss the possible relevance of our results to the
phase diagram of disordered integer quantum Hall systems.
\end{abstract}

\input epsf
The Hall conductance of Integer Quantum Hall systems is described
mathematically by the index of Fredholm operators. (For precise
definitions, see below). In this paper we investigate the phase
diagram of the Fredholm index for a few classes of operators. For
the algebra of bounded operators, little can be said beyond the
fact that the phase diagrams can be arbitrarily complicated. But
for the algebra of Toeplitz operators, and other related classes
of operators, we establish a kind of a Gibbs phase rule
\cite{gibbs}.  Typical of our results is the statement that if the
system is governed by two parameters, then one should expect jumps
by one at phase boundaries and jumps by up to 2 at triple points,
while jumps by more than two should never be observed.

We relate this behavior
to experimental results, conjectures and open problems
that arise in the context
of the Quantum Hall Effect (QHE) \cite{qhe}.

In Section 1 we define Fredholm operators and their indices, and
explore the different sorts of phase diagrams that can arise. In
Section 2 we recall how Fredholm indices  are related to the
conductance of Quantum Hall systems. In Section 3 we 
consider phase diagrams for general bounded operators. In
Section 4 we describe the phase diagram for linear combination of
shift operators, and in Section 5 we consider general Toeplitz operators.
In Section 6 we discuss the phase diagrams of soluble  models related
to the quantum Hall effect, and how they might be modified by disorder.
We also discuss the relevance of Toeplitz operators
to the Quantum Hall Effect and present some open problems.

\section{Fredholm indices}

\subsection{Basic notions}

The following is a brief description of Fredholm operators.
For more details, see \cite{atiyah,douglas,kato}.

\begin{Def}  A bounded operator $F$ on a separable Hilbert space is
{\em Fredholm} if there exists a bounded operator $B$ such that
$1-FB$ and $1-BF$ are compact. The Fredholm index is defined by
\begin{equation}
Index(F)= dim\, Ker(F)- dim\,Ker(F^\dagger).
\label{index}
\end{equation}
\end{Def}

The simplest example of a Fredholm operator with nonzero index is
the unilateral shift operator: Let $e_0, e_1, e_2, \ldots$ be the
canonical  basis for the Hilbert space $\ell^2( \mathbb{N})$, and
let the operator $a$ act by
\begin{equation}
a(e_n) =
\begin{cases}e_{n-1} & if \ n>0 \cr 0 & if \ n=0.\end{cases} \label{shift}
\end{equation}
The reason for denoting the unilateral shift operator by $a$ is its
similarity to the harmonic oscillator lowering operator. 
The adjoint of $a$ acts by
\begin{equation} a^\dagger(e_n) = e_{n+1}
\end{equation}
Since ${1}=a a^\dagger=
a^\dagger a + |e_0 \ket \bra e_0 |$, $a$ is Fredholm. The kernel
of $a$ is 1-dimensional and the kernel of $a^\dagger$ is
0-dimensional. Thus $Index\ (a)=1$ and $Index \ (a^\dagger)=-1$.

Although neither the dimension of $Ker\ F$ nor that of $Ker\, F^\dagger$
is stable under deformations of $F$, the index {\em is} stable. For any
compact operator $C$, for any bounded operator $B$, and for $\epsilon$
sufficiently small,
\cite{douglas,kato}:
\begin{equation} Index (F)= Index (F+\epsilon B+C).
\end{equation}

The following theorem is standard:
\begin{theorem} If $A_1, \ldots A_n$ are Fredholm operators, then
the product $A_1 A_2 \cdots A_n$ is also Fredholm, and
$Index(A_1\cdots A_n) = \sum_{i=1}^n Index(A_i)$.
\end{theorem}

If $F$ and $F'$ are Fredholm operators on the same
Hilbert space, then there is a continuous path of Fredholm
operators from $F$ to $F'$ if and only if $Index(F)=Index(F')$.
(By continuous, we mean relative to the operator norm).  Put
another way, the path components of $Fred(H)$, the space of
Fredholm operators on $H$, are indexed by the
integers.  The $n$-th path component is precisely the set of
Fredholm operators of index $n$  \cite{douglas}.

\subsection{Phase diagrams}
 Our main concern in this paper is the following problem:
Suppose one interpolates between Fredholm operators with different
indices. What can one say about the way the indices change?
Another  way of phrasing this is: What is the phase diagram of
Fredholm indices?

 The answer to this question depends on the
choice of the embedding space. In the space of bounded operators,
the ``phases''---each labeled by its index---are open sets. But
the boundary between phases, as we shall explain, is rather wild:
A point on the boundary of one phase is also on the boundary of
{\em every other} phase.  This behavior is difficult to visualize.

Another class of embedding spaces that we consider is associated
with Toeplitz operators with various regularity assumptions on a
class of functions. Here, at least if the functions are
sufficiently smooth,  the boundaries between phases have a simple
structure and the phase diagrams satisfy simple rules that have
the flavor of Gibbs' phase rule  \cite{gibbs}. 
Typical of our
results is the statement that under appropriate conditions, phases
whose indices differ by one have a common boundary whose
codimension is one, and phases whose indices differ by two meet on
a set of codimension two etc. Fig.\,\ref{shifts} is an example of
one of the phase diagrams we obtain.

\section{The Hall conductance as a Fredholm Index}

Theories of the quantum Hall effect are roughly of two kinds:
those that focus on the bulk of the Hall  and those that focus on
the edge  \cite{qhe}.  
It was pointed out by  \cite{frohlich} 
that the bulk-edge duality is an illustration of the {\em holographic
principle}.  In either approach, the quantized Hall conductance
can be related to a Fredholm index.

\subsection{Theories of the bulk} It is
common knowledge that the Hall conductance can be identified with
a Chern number  \cite{thouless}. 
For non-interacting electrons in
two dimensions, this result is a special case of the fact that the
Hall conductance is a Fredholm index. Since this is not common
knowledge, we recall how Chern numbers and Fredholm indices are
related.

For non-interacting electrons in two dimensions with the Fermi
energy in a gap, TKN$^2$, showed that the Hall conductance for
Landau Hamiltonians with {\em periodic} potential, is related to a
Chern number  \cite{tknn}. 
The (magnetic) Brillouin zone associated
with the periodicity plays an role in this theory. Because of
this, the interpretation of the Hall conductance as a Chern number
does not carry over to random or even quasi-periodic potentials
nor to ``irrational magnetic fields'', all of  which have no
(classical) Brillouin zone. Although the quantization of the Hall
conductance can be established in these cases by a limiting
argument  \cite{kunz,thouless}, 
the interpretation as a Chern
number does not survive.

J. Bellissard  \cite{bel}, 
in a work that had impact on
non-commutative geometry  \cite{connes,madore}, 
showed that the Hall conductance  with
{\em ergodic} potential, be it periodic, quasi-periodic or random, and
 real magnetic field, rational or not,
 is a Fredholm index.
This result was derived in  \cite{ass} 
without using non-commutative geometry.

 More precisely,  consider the (infinite dimensional) spectral projection $P$
 on the states below the Fermi energy
$E_F$ for the one particle Hamiltonian in the plane.  Let $U$ be
the multiplication operator $e^{i\theta}$, where $\theta$ is the
usual polar angle in the plane. $U$ is a singular gauge
transformation that introduces an Aharonov-Bohm flux tube at the
origin of the Euclidean plane. The Hall conductance is the
Fredholm index
  of $PUP$ thought of as an operator on the
range of $P$  \cite{foot1}. 
Since the Fredholm index does not need a
Brillouin zone, this approach offers a natural framework that
accounts  for the quantization and stability of the Hall
conductance.

\subsection{Theories of the edge}\label{edge}

Finite quantum Hall systems have chiral edge currents
\cite{halperin,walcher}. 
Consider the case that the boundary is a
circle of circumference $L$. The dispersion relation of the edge
states is approximately linear in a small neighborhood of the
Fermi energy and the Hamiltonian for a single edge channel, with
velocity $v_F$, is
\begin{equation}
H= -i\frac{v_F}{L}\partial_\theta
\end{equation}
 Now, the projection $P$ is associated with the occupied edge
 states, $e^{-im\theta}$ with  $m \ge m_0$.
 Introducing a flux tube  into the system is associated with the
 unitary $U=e^{i\theta}$ and sends $H\to UHU^\dagger$. This
 leads to the spectral flow of the edge states. $PUP$ is the unilateral shift
operator $a$ and the number of edge states
 that cross the Fermi energy is $Index\, PUP=1$. By an argument of
Halperin  \cite{halperin} 
this is also the Hall conductance.

An extension of this idea to Harper models with an edge is
described in  \cite{SB}.

\section{The phase diagram for bounded operators}

We begin with the space of bounded operators with
the topology defined by the operator norm, and we wish
to understand the phase diagram of a generic family of
such operators. As we
shall explain, the phase diagram in the entire space is quite wild:
Any point on the boundary of the ``index = $k$'' phase
is also on the boundary of every other phase.

To understand this bizarre behavior, recall that the zero operator
(which is {\em not} a Fredholm operator) is on the boundary of every
phase: Zero is the limit, as $\varepsilon \to 0$, of
$\varepsilon a^n$, with $a$ of Eq.~(\ref{shift}), for
any $n$. The point of the theorem is that similar behavior occurs at
all boundary points.

\begin{theorem} Let $U_n$ be the set of Fredholm operators of index $n$.
Every point on the boundary of $U_n$ is also on the boundary of
$U_m$, for every integer $m$.
\end{theorem}

Proof: Let $A$ be a (not Fredholm) operator on the boundary of $U_n$.  Given $\epsilon >0$,
we must find an operator in $U_m$ within a distance $\epsilon$ of $A$.

Suppose that the kernel and cokernel of $A$ are infinite
dimensional, and that there is a gap in the spectrum of $A^\dagger
A$ at zero.  (If this is not the case, we may perturb $A$ by an
arbitrarily small amount to make it so).  Now let $B$ be a unitary
map from the kernel of $A$ to the cokernel.   Let $P,\ (P')$ be
the orthogonal projection onto $ker(A), \ (coker(A))$, and let $a$
be a shift operator on $ker(A)$. For each $m \ge 0$,
$A(\epsilon)=A + \epsilon B a^m P$ has a bounded right inverse
\begin{equation}
A^\dagger {1\over P'+ AA^\dagger} P'_\perp + {1\over\epsilon}
(a^\dagger)^m B^\dagger P'.
\end{equation}
It follows that the cokernel of
$A(\epsilon)$ is empty. It is easy to see that the kernel of
$A(\epsilon)$ is $m$ dimensional hence $Index(A(\epsilon))=m$.
Similarly, $A + \epsilon B (a^\dagger)^{m} P$ has index $-m$.
\hfill$\qed$

\section{Linear combinations of shifts}

In this section and the next we show that there are interesting
and simple  ``generic'' phase diagrams of Fredholm indices in some
finite dimensional spaces, and in some infinite-dimensional spaces
with sufficiently fine topologies.  We shall also see also how
control is lost as the space is enlarged and the topology is
coarsened.

\subsection{Shift by one}

We begin by considering linear combinations of the shift operator $a$ and the identity
operator 1.  That is, we consider the operator
$$ A = c_1 a + c_0 $$
where $c_1$ and $c_0$ are constants.

\begin{theorem}
\label{0-1}
If $|c_1| \ne |c_0|$, then $A$ is Fredholm.
The index of $A$ is 1 if $|c_1|>|c_0|$ and zero if $|c_1|<|c_0|$.
If $|c_1| = |c_0|$, then $A$ is not Fredholm.
\end{theorem}

\nd Proof: First suppose $|c_0| > |c_1|$.  Then $A$ is invertible:
$$ A^{-1} = c_0^{-1} (1 + (c_1/c_0) a)^{-1} = \sum_{n=0}^\infty {(-1)^n c_1^n
\over c_0^{n+1}} a^n,
$$
as the sum converges absolutely.  Thus $A$ has neither kernel nor cokernel, and has
index zero.

If $|c_1| > |c_0|$, then the kernel of $A$ is 1-dimensional, namely all multiples of
$|\psi \ket = \sum_{n=0}^\infty z_0^n e_n$, where $z_0 = -c_0/c_1$. Notice how the norm
of $|\psi \ket$ goes to infinity as $|z_0| \to 1$.  However, $A^\dagger$ has no kernel,
since for any unit vector $|\phi \ket$,
$\|A^\dagger | \phi \ket\| = \|\bar c_1 a^\dagger |\phi \ket + \bar c_0 |\phi \ket\|
\ge   \|\bar c_1 a^\dagger |\phi \ket \| - \|\bar c_0 |\phi \ket\|
= |c_1| - |c_0|$.  Thus the index of $A$ is 1.

If $|c_1| = |c_0| $, then $A$ is at the boundary between index 1
and index 0, and so cannot be Fredholm.\hfill$\qed$

\subsection{Finite linear combinations of shifts}

Next we consider linear combinations of $1, a, a^2, \ldots$ up to some fixed $a^n$.
That is, we consider operators of the form
\begin{equation} A = c_n a^n + c_{n-1} a^{n-1} + \cdots + c_0.
\label{A}
\end{equation}
This is closely related to the polynomial
\begin{equation} p(z) = c_n z^n + \cdots + c_0.
\label{p}
\end{equation}

\begin{theorem}
If none of the roots of $p$ lie on the unit circle, then $A$ is Fredholm, and the
index of $A$ equals the number of roots of $p$ {\em inside} the unit circle, counted with
multiplicity.  If any of
the roots of $p$ lie {\em on} the unit circle, then $A$ is not Fredholm.
\end{theorem}

\nd Proof: The polynomial $p(z)$ factorizes as $p(z) = c_k \prod_{i=1}^k (z-\zeta_i)$, where $k$
is the degree of $p$ (typically $k=n$, but it may happen that $c_n=0$).  But then
$A = c_k \prod_{i=1}^k (a - \zeta_i)$. If none of the roots $\zeta_i$ lie on the unit circle,
then each term in the product is Fredholm, so the product is Fredholm, and the index of
the product is the sum of the indices of the factors.  By Theorem \ref{0-1}, this exactly
equals the number of roots $\zeta_i$ inside the unit circle.

If any of the roots lie on the unit circle, then a small
perturbation can push those roots in or out, yielding Fredholm
operators with different indices. This borderline operator
therefore cannot be Fredholm.\hfill$\qed$

The last theorem easily generalizes to linear combination of
left-shifts and right-shifts.  The index of an operator
\begin{equation} A = c_n a^n + \cdots + c_1 a + c_0 + c_{-1} a^\dagger + \cdots + c_{-m} (a^{\dagger})^m
\end{equation}
equals the number of roots of
\begin{equation} p(z) = \sum_{i=-m}^n c_i z^i
\end{equation}
inside the unit circle, minus the degree of the pole at $z=0$ (that is
$m$, unless $c_{-m}=0$).  This follow from the fact that
\begin{equation} A = (\sum_{i=-m}^n c_i a^{i+m}) (a^\dagger)^m.
\end{equation}

Since there is no qualitative difference between combinations of
left-shifts and combinations of both left- and right-shifts, we
restrict our attention to left-shifts only, and consider families of
operators of the form (\ref{A}).

\begin{theorem}\label{finite-combinations}
In the space of complex linear combinations of 1, $a$, \ldots, $a^n$,
almost every operator is Fredholm.  For every $k \le n$, the points
where the index can jump by $k$ (by which we mean the common
boundaries of regions of Fredholm operators whose indices differ by
$k$) is a set of real codimension $k$.

In the space of real linear combinations of 1, $a$, \ldots, $a^n$,
almost every operator is Fredholm.  For every $k \le n$, the points
where the index jumps by $k$ is a stratified space, the largest
stratum of which has real codimension $\lfloor (k+1)/2 \rfloor $,
where $ \lfloor x \rfloor$ denotes the integer part of $x$.
\end{theorem}

\nd Proof: Our parameter space is the space of coefficients $c_i$, or
equivalently the space of polynomials of degree $\le n$. This is
either $\R^{n+1}$ or $\C^{n+1}$, depending on whether we allow real or
complex coefficients.  In either case, the set $U_k$ of Fredholm
operators of index $k$ is identical to the set of polynomials with $k$
roots inside the unit circle and the remaining $n-k$ roots outside (if
$c_n=0$, we say there is a root at infinity; if $c_n=c_{n-1}=0$, there
is a double root at infinity, and so on. Counting these roots at
infinity, there are always exactly $n$ roots in all.) The boundary of
$U_k$ is the set of polynomials with at most $k$ roots inside the unit
circle, at most $n-k$ outside the unit circle, and at least one root
on the unit circle. (Strictly speaking, the zero polynomial is also on
this boundary.  This is of such high codimension that it has no effect
on the phase portrait we are developing.).  We consider the common
boundary of $U_k$ and $U_{k'}$.  If $k<k'$, a nonvanishing polynomial
is on the boundary of both $U_k$ and $U_{k'}$ if it has at most $k$
roots inside the unit circle and at most $n-k'$ roots outside.  It
must therefore have at least $k'-k$ roots on the unit circle.

If we are working with complex coefficients, this is a set of
codimension $k'-k$. The roots themselves, together with an overall
scale $c_n$, can be used to parametrize the space of polynomials.  For
each root, being on the unit circle is codimension 1, while being
inside or outside are open conditions.  Since the roots are
independent, placing $k'-k$ roots on the unit circle is codimension
$k'-k$.

If we are working with real coefficients, the roots are not
independent, as non-real roots come in complex conjugate pairs.  Thus,
the common boundary of $U_k$ and $U_{k'}$ breaks into several strata,
depending on how many real roots and how many complex conjugate pairs
lie on the unit circle.  If $k'-k$ is even, the biggest stratum
consists of having $(k'-k)/2$ pairs, and has codimension $(k'-k)/2$.
If $k'-k$ is odd, the biggest stratum consists of having $(k'-k-1)/2$
pairs and one real root on the unit circle, and has codimension
$(k'+1-k)/2$.  \hfill\qed

\begin{figure}
\vskip -1in \centerline{\hskip -1.3in\epsfxsize=3.0truein
\epsfbox{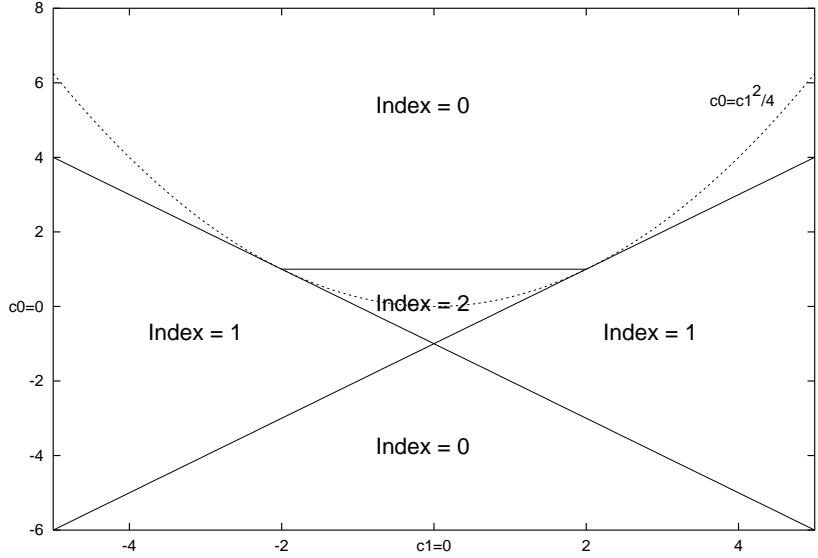}} \caption{A phase diagram for the Fredholm
index of $F=a^2 + c_1 a + c_0$.} \label{shifts}
\end{figure}

Theorem \ref{finite-combinations} is illustrated in
Figure~\ref{shifts}, where the phase portrait is shown for $n=2$
with real coefficients, with $c_2$ fixed to equal 1. The points
above the parabola $c_0 = c_1^2/4$ have complex conjugate roots,
while points below have real roots. Notice that the transition
from index 2 to index 0 occurs at an isolated point when the roots
are real, but on an interval when the roots come in
complex-conjugate pairs.

It is clear that an almost identical theorem applies to linear
combinations of left-shifts up to $a^n$ and right-shifts up to
$(a^\dagger)^m$.  The results are essentially independent of $n$ and
$m$ (their only effect being to limit the size of possible jumps to
$n+m$). We can therefore extend the results to the space of all
(finite) linear combinations of left- and right-shifts, which is
topologized as the union over all $n$ and $m$ of the spaces considered
above. Our result, restated for that space, is

\begin{theorem}\label{limit-space}
In the space of finite complex linear combinations of left- and
right-shifts of arbitrary degree, almost every operator is Fredholm.
For every integer $k \ge 1$, the points where the index can jump by
$k$ (by which we mean the common boundaries of regions of Fredholm
operators whose indices differ by $k$) is a set of real codimension
$k$.

If we restrict the coefficients to be real, then, for every $k \le n$,
the points where the index jumps by $k$ is a stratified space, the
largest stratum of which has real codimension $\lfloor (k+1)/2 \rfloor
$.
\end{theorem}

\section{Toeplitz operators}\label{toeplitz}

Although Theorem \ref{limit-space} refers to an infinite-dimensional
space, this space is still extremely small -- each point is a {\em
finite} linear combination of shifts. In this section we consider {\em
infinite} linear combinations of shifts. This is equivalent to
studying Toeplitz operators.

\begin{Def} The {\em Hardy space} $H$ is the subspace of $L^2(S^1)$
consisting of functions whose Fourier transforms have no negative
frequency terms.  Equivalently, if we give $L^2(S^1)$ a basis of
Fourier modes $e_n = e^{in\theta}$, where the integer $n$ ranges from
$-\infty$ to $\infty$, then $H$ is the closed linear span of $e_0,
e_1, e_2, \ldots$.
\end{Def}

We think of $S^1$ as sitting in the complex plane, with
$z=e^{i\theta}$. Now let $f(z)$ be a bounded, measurable function on
$S^1$, and let $P$ be the orthogonal projection from $L^2(S^1)$ to
$H$.  If $| \psi \ket \in H$, then $|f \psi\ket$ (pointwise product)
is in $L^2(S^1)$, and $P|f \psi\ket \in H$.  We define the operator
$T_f$ by
\begin{equation} T_f |\psi\ket = P|f \psi\ket. \label{Toep}
\end{equation}

\begin{Def}
An operator of the form (\ref{Toep}) is called a {\em Toeplitz
operator}. We call a Toeplitz operator $T_f$ continuous if the
underlying function $f$ is continuous, and apply the terms
``differentiable", ``smooth" and ``analytic" similarly.
\end{Def}

\nd {\bf Remark}: Toeplitz operators can be represented by
semi-infinite matrices that have constant entries on diagonals, and
the various classes we have defined correspond to the decay away from
the main diagonal.

Notice that \begin{equation} T_{e_m} e_n = \begin{cases} e_{n+m} & if
\ n+m \ge 0 \cr 0 & otherwise\end{cases} \end{equation} so $T_{e_m}$
is simply a shift by $m$, a right shift if $m>0$ and a left-shift if
$m<0$. All our results about shifts can therefore be understood in the
context of Toeplitz operators. Theorem \ref{finite-combinations}
refers to operators $T_f$, where $f$ is a polynomial in $z^{-1}$ of
limited degree. Theorem \ref{limit-space} considers polynomials or
arbitrary degree in $z$ and $z^{-1}$.  We will see that the results
carry over to analytic functions on an annulus around $S^1$, and to a
lesser extent to $C^k$ Toeplitz operators, but with results that
weaken as $k$ is decreased.

Here are some standard results about Toeplitz operators.
 For details, see  \cite{douglas}.

\begin{theorem} A $C^1$ Toeplitz operator $T_f$ is Fredholm if
and only if $f$ is everywhere nonzero on the unit circle.  In that
case the index of $T_f$ is minus the winding number of $f$ around the
origin, namely
\begin{equation} Index(T_f) = -Winding(f) = {-1\over 2 \pi i} \int_{S^1} {df \over f}, \label{wind}
\end{equation}
\end{theorem}

Given the first half of the theorem, the equality of index and
winding number is easy to understand.  We simply deform $f$ to a
function of the form $f(z)=z^n$, while keeping $f$ nonzero on all
of $S^1$ throughout the deformation (this is always possible, see
e.g.  \cite{gp}). 
In the process of deformation, neither the index
of $T_f$ nor the winding number of $f$ can change, as they are
topological invariants.  Since the winding number of $z^n$ is $n$,
and since $T_{z^n} = (a^\dagger)^n$ (if $n\ge 0$, $a^{-n}$
otherwise), which has index $-n$, the result follows.

We now consider functions $f$ on $S^1$ that can be analytically
continued (without singularities) to an annulus $r_0 \le |z| \le
r_1$, where the radii $r_0 < 1$ and $r_1>1$ are fixed. This is
equivalent to requiring that the Fourier coefficients $\hat f_n$
decay exponentially fast, i.e. that the sum
\begin{equation}
\sum_{n=-\infty}^\infty |\hat f_n |(r_0^n + r_1^n)
\end{equation}
converges. For now we do not impose
any reality constraints or other symmetries on the coefficients
$\hat f_n$.  This space of functions is a Banach space, with norm
given by the sup norm on the annulus.  This norm is stronger than
any Sobolev norm on the circle itself.

The analysis of the corresponding Toeplitz operators is
straightforward and similar to the proof of Theorem
\ref{finite-combinations}.  Since $f$ has no poles in the annulus, we
just have to keep track of the zeroes of $f$.  For the index of $T_f$
to change, a zero of $f$ must cross the unit circle.  For the index to
jump from $k$ to $k'$, $|k-k'|$ zeroes must cross simultaneously.  In
the absence of symmetry, the locations of the zeroes are independent
and can be freely varied, so this is a codimension-$|k-k'|$ event.

If we impose a reality condition: $f(\bar z) = \overline{f(z)}$, then
zeroes appear only on the real axis or in complex conjugate pairs. In
that case, changing the index by 2 is merely a codimension-1 event.
Combining these observations we obtain

\begin{theorem}\label{analytic}
In the space of Toeplitz operators that are analytic in a (fixed)
annulus containing $S^1$, almost every operator is Fredholm.  For
every integer $k \ge 1$, the points where the index can jump by
$k$ is a set of real codimension $k$.

If we impose a reality condition $f(\bar z)=\overline{f(z)}$ then,
for every $k \le n$, the points where
the index jumps by $k$ is a stratified space,
the largest stratum of which has real codimension $\lfloor (k+1)/2 \rfloor $.
\end{theorem}

Finally we consider Toeplitz operators that are not necessarily
analytic, but are merely $\ell$ times differentiable, and we use the
$C^\ell$ norm.  Our result is

\begin{theorem}\label{C^ell}
In the space of Toeplitz $C^\ell$ operators, almost every operator is
Fredholm.  For every integer $k$ with $1 \le k \le 2\ell+1$, the
points where the index can jump by $k$ is a set of real codimension
$k$.  For every integer $k \ge 2\ell+1$, the points where the index
can jump by $k$ is a set of real codimension $2\ell+1$.
\end{theorem}

In other words, our familiar results hold up to codimension $2\ell+1$,
at which point we lose all control of the change in index.

\smallskip

\nd Proof: As long as $f$ is everywhere nonzero, $T_f$ is Fredholm.
To get a change in index, therefore, we need one or more points where
$f$, and possibly some derivatives of $f$ with respect to $\theta$,
vanish.  Suppose then that for some angle $\theta_0$,
$f(\theta_0)=f'(\theta_0)=\cdots=f^{(n-1)}(\theta_0)=0$ for some $n
\le \ell$, but that the $n$-th derivative $f^{(n)}(\theta_0) \ne
0$. This is a codimension $2n-1$ event, since we are setting the real
and imaginary parts of $n$ variables to zero, but have a 1-parameter
choice of points where this can occur.  Without loss of generality, we
suppose that this $n$-th derivative is real and positive.  By making a
$C^\ell$-small perturbation of $f$, we can make the value of $f$
highly oscillatory near $\theta_0$, thereby wrapping around the origin
a number of times.  However, since a $C^\ell$-small perturbation does
not change the $n$-th derivative by much, the sign of the real part of
$f$ can change at most $n$ times near $\theta_0$, so the argument of
$f$ can only increase or decrease by $n\pi$ or less.  The difference
between these two extremes is $2n\pi$, or a change in winding number
of $n$.

To change the index by an integer $m$, therefore, we must have the
function vanish to various orders at several points, with the sum of
the orders of vanishing adding to $m$. The generic event is for $f$
(but not $f'$) to vanish at $m$ different points -- this is a
codimension $m$ event, analogous to having $m$ zeroes of a polynomial
cross the unit circle simultaneously at $m$ different points. All
other scenarios have higher codimension and are analogous to having 2
or more zeroes of the $m$ zeroes crossing the unit circle at the same
point.

The situation is different, however, when the function $f$ and the
first $\ell$ derivatives all vanish at a point $\theta_0$. Then
the higher-order derivatives are not protected from $C^\ell$-small
perturbations and, by making such a perturbation, we can change
$f$ into a function that is identically zero on a small
neighborhood of $\theta=\theta_0$. By making a further small
perturbation, we can make $f$ wrap around the origin as many times
as we like near $\theta=\theta_0$. More specifically, if $f$ is
zero on an interval of size $\delta$, then, for small $\epsilon$,
$\tilde f(\theta) = f(\theta) + \epsilon e^{iN\theta}$ will wrap
around the origin approximately $N\delta/2\pi$ times near
$\theta_0$.  By picking $N$ as large (positive or negative) as we
wish, we can obtain arbitrarily positive or negative indices.  As
long as we take $\epsilon \ll N^{-\ell}$, this perturbation will
remain small in the $C^\ell$ norm.\hfill\qed

The results of this section can be extended, with minor modifications,
to the algebra of matrix valued Toeplitz operators \cite{douglas}
where the index is related to the winding of the determinant of a matrix.

\section{Quantum Hall systems}
\subsection{Phase diagrams of soluble models}

\begin{figure}
 \centerline{
 \epsfbox{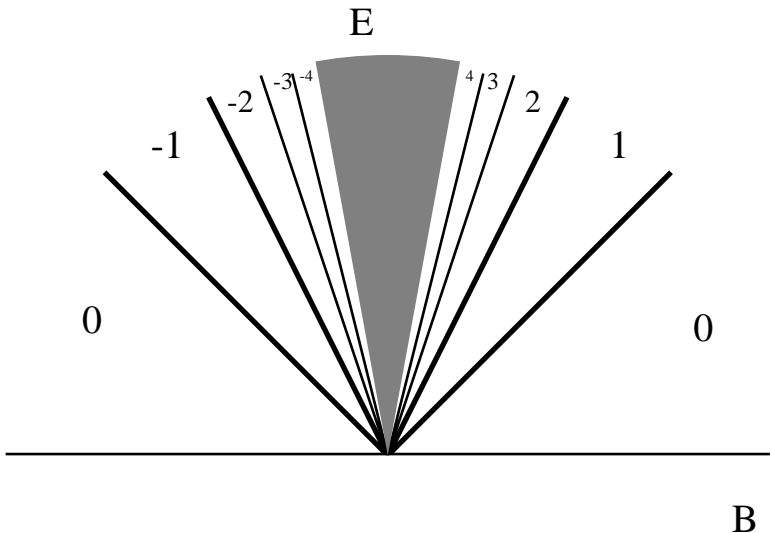}
 }
\caption{The phase diagram for the Landau Hamiltonian in the
Euclidean plane. The shaded wedge contains infinitely many,
thinner and thinner, wedges, with indices that go to $\pm\infty$
and accumulate at the $B$ axis. }\label{landau}
\end{figure}

\begin{figure}
 \centerline{\epsfxsize=2.5truein
 \epsfbox{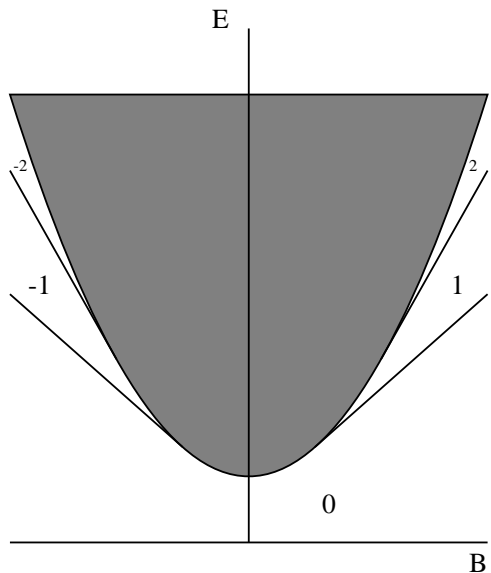}}
\caption{The phase diagram for the Landau Hamiltonian in the
hyperbolic plane. In the shaded parabolic region the operator is
not Fredholm and the index is nor defined. }\label{hyperbolic}
\end{figure}

\begin{figure}
\vskip -0.8 in
\centerline{\hskip -1.2in \epsfxsize=4.5truein\epsfbox{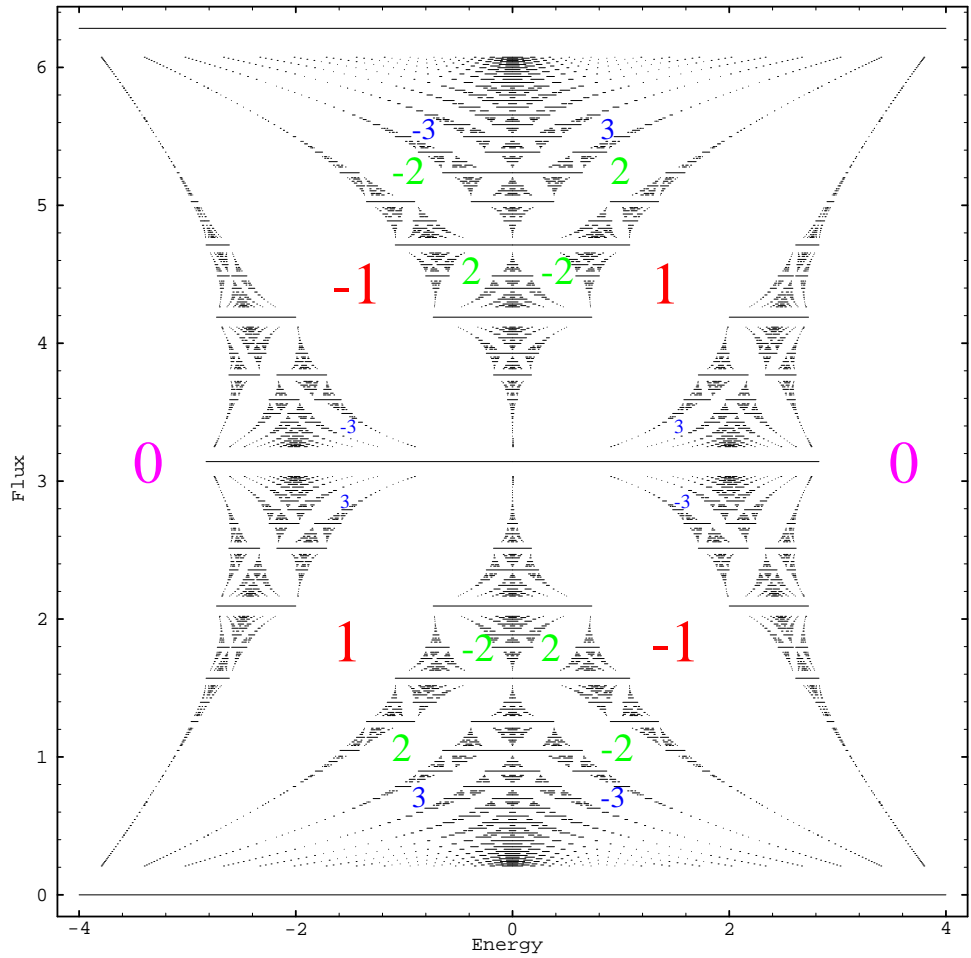} }
\vskip -0.4 in
\caption{The phase diagram for the
Harper model associated with tight binding model on a square
lattice the plane. Every point on the boundary between two phases
appears to be a point of accumulation of infinitely many phases.
Figure taken, with permission, from \cite{ks}.}
\label{harper}
\end{figure}

Phase diagrams of quantum Hall system describe the dependence of
the Hall conductance on parameters such as the magnetic field $B$
and the Fermi energy $E$. There are three idealized models where
the phase diagram can be computed explicitly: The Landau
Hamiltonian in the Euclidean plane, whose phase diagram is shown
in Fig.\,\ref{landau}; The Landau Hamiltonian for the hyperbolic
plane, whose  phase diagram is shown in Fig.\,\ref{hyperbolic} and
Harper models in the plane  \cite{hofstadter,tknn}, 
whose phase diagram  is associated with the Hofstadter butterfly, shown in
Fig.\,\ref{harper} for the case of a tight binding model on a
square lattice.

These are not models of Toeplitz operators, and none of these models
is generic, especially insofar as all of
them have symmetries.  However, we consider the extent to which
they follow the generic phase rules of (smooth, complex) Toeplitz operators
anyway. Where these rules are
not followed, we consider how a small generic perturbation might
restore the rules.

 The phase
diagram for the Euclidean plane, Fig. \ref{landau} satisfies
the generic phase rules away from the line $B=0$.
On the line $B=0$, however, the index takes an infinitely large
jump, while at the origin infinitely many phases meet.
Both are forbidden by the phase rules.

The phase diagram in the hyperbolic plane, Fig. \ref{hyperbolic},
satisfies the generic phase rules outside
the shaded parabolic region.  In the shaded region, the operator
is not Fredholm  and the index is not defined. This is contrary to
the phase rules since not being Fredholm is expected to be
a codimension 1 event.

The phase diagram of the Harper model, Fig.~\ref{harper}, is in
serious conflict with the phase rule for (smooth, complex) Toeplitz
operators: It is known  \cite{last}, 
that for a full measure of values
of the magnetic field (irrational, of course), the spectrum is a
Cantor set. Since the boundary between phases is contained in the
spectrum, this suggests that any point on the boundary between any two
phases can also be on the boundary between infinitely many other
phases. This is the sort of behavior we observed for bounded operators
with no restrictions.  However, even in this wildness there is some
regularity.  For example, the center of the figure is on the boundary
of all phases with {\em odd} indices while Theorem 2 allows for
even indices as well.

\begin{Remark}  To see how Fig.~\ref{harper} is
obtained, we recall that for a tight-binding model with flux
$\frac p q$ through a unit cell, the Hall conductance, $\sigma_j$
associated with the j-th gap, (provided all gaps below it are
open) satisfies the Diophantine equation  \cite{tknn,daz}
\begin{equation}\label{dio}
p \sigma_j =j \, mod\, q.
\end{equation}
A similar equation
holds for gaps counted from above. In the Harper model it is known
\cite{mouche} 
that all gaps except possibly for the central gap,
are open.
\end{Remark}

\begin{figure}
 \centerline{\epsfxsize=2.5truein
 \epsfbox{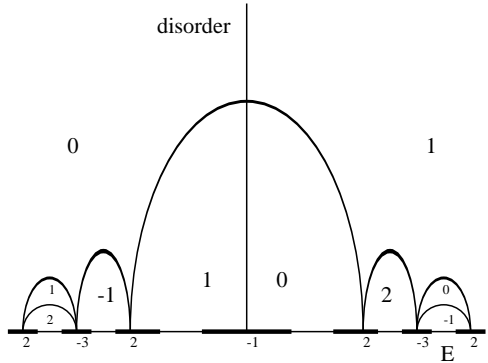}}
\caption{The phase diagram for the Hall conductance of a split
Landau level in Harper model with disorder after
\cite{tan}.}
\label{tanf}
\end{figure}

Finally, consider the phase diagram of the Harper model with a
disordered potential. This is not soluble in the same sense that
the previous models are, but there are numerical results for it.
Fig.\,\ref{tanf}, which we borrowed from  \cite{tan}, 
shows the phase diagram for a split Landau level in the Harper model with
disorder. More precisely, the diagram describes a Harper model
with fractional flux $\frac 8 5$ through a unit cell.

Without disorder the conductance $\sigma$ of each  isolated band
satisfies the Diophantine equation similar to Eq. \ref{dio},
except that for a split Landau band $p$ and $q$ are interchanged.
For flux $\frac 8 5$ the Diophantine equation fixes the
conductances $(2,-3,2)$ of the bands at the flanks and $-1$ at the
center. Zero disorder is, of course, not generic, and, indeed,
there are bands on the $E$ axis where the index is not defined,
something that the phase rules for Toeplitz forbid. Under
perturbation the diagram should might so that these bands where
the index is not defined disappear. This is indeed the
case. The diagram in  \cite{tan} 
is obtained by drawing $n$ lines
emanating from each band where $n$ is its Hall conductance.

In summary, the wild character of the phase diagram of the Harper model
is tamed by disorder and one finds, remarkably, a phase
diagram compatible with the phase rules for Toeplitz operators.

\subsection{Perturbations of Landau Hamiltonians}

Motivated by the effect of disorder on the Harper model phase
portrait, we next consider the effect of perturbations on
the phase portraits of Landau Hamiltonians.  
Such perturbations will modify the phase diagram near phase
boundaries. As a consequence one expects a phase diagram to be
qualitatively modified near points of accumulation of phases, even
if the perturbation is small.

Figures 2 and 3 satisfy the phase rules in the region of
large magnetic fields, but fail to do so for small
magnetic fields. We now examine how the two figures might be
modified to satisfy the phase rules everywhere.

\begin{figure}
  \centerline{
  \epsfxsize=2.5truein
\epsfbox{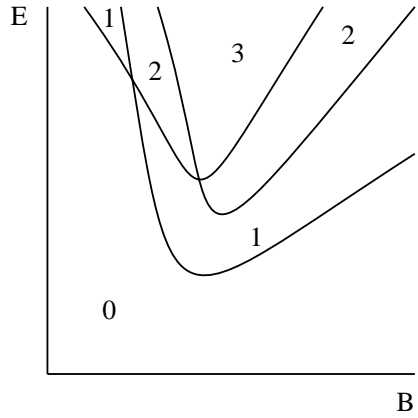}}\caption{A phase diagram that satisfies the
phase rules of Toeplitz operators and is a perturbation of the
phase diagram of Landau Hamiltonian in the plane.
}\label{landau+toeplitz}
\end{figure}

The phase diagram of the Landau Hamiltonian in the plane, Fig.
\ref{landau}  will be significantly modified near the line $B=0$
which,  by symmetry, must lie in a region with index $0$. A
schematic phase diagram that is generic and close to the Landau
phase diagram is shown in Fig. \ref{landau+toeplitz}.

\begin{figure}
  \centerline{\epsfysize=2.5truein
\epsfbox{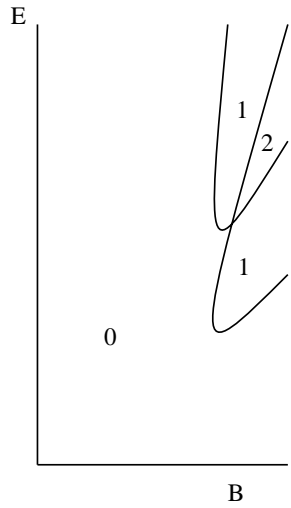}}\caption{A phase diagram that satisfies
the rules of Toeplitz operators and is a perturbation of the phase
diagram of Landau Hamiltonian in the hyperbolic plane.
}\label{hyper-toeplitz}
\end{figure}

 The phase diagram in Fig.
\ref{hyperbolic} has a region of full measure, the shaded
parabola, where the operator is not Fredholm. This is non-generic,
and unstable. A perturbation might produce a phase diagram like
\ref{hyper-toeplitz}. Note that the two perturbed diagrams, Fig.
\ref{landau+toeplitz} and \ref{hyper-toeplitz} are topologically
identical.

How do the phase diagrams, Figs. \ref{landau+toeplitz} and
\ref{hyper-toeplitz}, compare with what one finds in experiments on
the quantum Hall effect? For large magnetic fields one finds phase
diagrams that resemble both Figs. \ref{landau} and
\ref{landau+toeplitz} and  satisfy the phase rules. For weak
magnetic fields one observes a transition to an insulating phase.
The emergence of an insulating phase (with index 0) for small
magnetic fields is in agreement with the phase rule and
Fig.~\ref{landau+toeplitz}. However, some experiments
\cite{shahar} and numerical simulations \cite{niu} 
have been interpreted as giving evidence to  direct
transitions from a Hall conductance of 2 and 3 to the insulating
phase. Taken literally, such transitions would violate the phase rule.
However, these results may merely indicate that, for $B$ small, the
phase boundaries of Fig. \ref{landau+toeplitz} are too closely 
spaced to be distinguished numerically and experimentally.

\subsection{Toeplitz operators}

The main gap in our analysis is that we have not established a direct
relation between the algebra of Toeplitz operators, where
our phase rules are proven, and the class of operators
relevant to (disordered) quantum Hall systems.

At the minimum, Toeplitz operators serve as a natural mathematical
laboratory. However, there is a more direct justification for
considering Toeplitz operators. The most elementary paradigm for a
quantum Hall system is the Landau Hamiltonian, in which case one
has:

\begin{theorem} Let $P$ be a projection on the lowest Landau level
in $\real^2$, and let $U$ be the gauge transformation associated with an
Aharonov-Bohm flux tube at the origin.  Then $PUP$, acting on the
range of $P$, differs from a Toeplitz operator by a compact operator.
\end{theorem}
\nd Proof: A basis for the lowest Landau level is
\begin{equation}
|n\ket={1\over \sqrt{\pi\, n!}}\, z^n\,e^{-|z|^2/2},\quad n\ge 0.
\end{equation} As a consequence
\begin{equation}
\bra n|U|m\ket=\delta_{n,m+1}\, {(m+1/2)!\over
m!\sqrt{m+1}}\approx \delta_{n,m+1}\left(1-{1\over 8m}\right).
\hfil\qed \end{equation}

In this case, a compact perturbation of $PUP$ is not only a
Toeplitz operator; it is a simple shift.
However, if the flux tube is placed at a different point, or if the magnetic
field is spread out over a finite region, then we obtain
a more general Toeplitz operator.
If $P$ is a projection on a higher Landau level, the same results
hold but the calculation is more involved. If $P$ is a projection
onto multiple Landau levels, then $PUP$ is a compact perturbation
of a direct sum of Toeplitz operators, one for each Landau level.

This is not to say that Toeplitz operators apply directly to all
systems, only that they apply to many. There are basic models
where $PUP$ fails to be
Toeplitz. Indeed, an elementary model for localization is a random
multiplication operator, i.e. $H=V_\omega$ on $\ell(\mathbb{Z}^d)$.
This is a caricature of strong disorder.  The eigenfunctions are now
concentrated at lattice points. The projection $P$ (below a Fermi
energy) is 
\begin{equation}
P=\sum |n\rangle\langle n|,
\end{equation}
where the sum is over a
random set of lattice points with, $V_\omega(n)<E$, in
$\mathbb{Z}^d$. $PUP$ is now a multiplication by a phase. It is an
invertible operator and has Fredholm index zero. It is, however,
not Toeplitz.

\subsection{Open problems}

It is tempting to directly study the index of $PUP$, for spectral projections
$P$ and unitary operators $U$, rather than rely on generic results
based on Toeplitz operators.  There are, however, several technical
obstacles.
 The first is that $PUP$ is thought of as acting on $Range \ P$, which
is a Hilbert space in its own right. This
means that a deformation of the parameters of the system
leads to a deformation of the space
$Range\ P$. In contrast, our strategy so far is formulated on a
fixed space. The second obstacle is that our results depend on
continuity properties while spectral projections tend to have bad
continuity properties that come from a discontinuity at the Fermi
energy.

 To overcome the first problem one can replace $PUP$ by an operator $F$
 defined on the entire Hilbert
space with coinciding index. There is large arbitrariness in
choosing $F$, but a natural choice is:
\begin{equation}\label{f}
F=PUP +P_\perp=PUP +P_\perp^2=1+P(U-1)P,\end{equation}
where $P_\perp=1-P$.

To overcome the second problem one may want to replace $P$ by a
Fermi function. That is, replace $P$  by a smooth version
\begin{equation}\label{temp} P(\beta,B,E_F)=
\frac{1}{\exp (\beta (H(B)-E_F)) +1}.
\end{equation}
In that case, however, $P^2$ is no longer equal to $P$, and the different
expressions for $F$ in equation (\ref{f}) are no longer equivalent.
For each choice, it would be interesting to derive a phase portrait
for index$(F)$ as the temperature, Fermi energy, magnetic field and
degree of randomness are varied.

\subsection{Concluding remark}

In this paper we explored what can be said about generic phase
diagrams of indices of Fredholm operators. We did not use the fact
that the Fredholm operators relevant to the quantum Hall effect
are of the form $PUP$, with $P$ a spectral projection of an
ergodic Schr\"odinger operator. Rather, we considered the index of
several natural classes (and algebras) of operators. The weakness
of this strategy is that we can not say much that is definitive
about quantum Hall systems. In its defense, we recall that
replacing the particular by the generic proved to be useful in
quantum physics in the hands of Wigner, von Neuman and Dyson
\cite{mehta,wvn,dyson}. 
Whether it will turn out to be useful for
quantum Hall effect remains to be seen.

\section*{Acknowledgments} We thank A. Kamenev
for drawing our attention to ref  \cite{sheng}, 
and E. Park, M. Reznikov, 
H. Schultz-Baldes and E. Shimshoni for useful discussions. This
research was supported in part by the Israel Science Foundation,
the Fund for Promotion of Research at the Technion, the DFG, the
National Science Foundation and the Texas Advanced Research
Program.


\begin{thebibliography}{article}
\bibitem[14]{foot1} $PUP$ is  Fredholm provided the integral
kernel of the projection, $p(\vec x,\vec y;E)=\langle \vec x
|P(E)|\vec y\rangle$ has good decay properties as $|\vec x-\vec
y|$ gets large. Such decay is indeed guaranteed by the theory of
localization whenever the Fermi energy is in the localized regime
\cite{aizenman}.
\end{thebibliography}
\end{document}